\documentclass[a4paper,twocolumn,times,astrosym,iop,useAMS,usenatbib]{aastex63}
\usepackage{amsmath}

\received{13th September 2020}
\revised{4th January 2020}
\accepted{5th January 2020}
\submitjournal{ApJ}

\shorttitle{[CII] emission in SL2S~0217}
\shortauthors{Rybak et al.}

\begin{document}


\title{Ultra-faint [\ion{C}{2}] emission in a redshift = 2 gravitationally-lensed metal-poor dwarf galaxy}


\correspondingauthor{Matus Rybak}
\email{mrybak@strw.leidenuniv.nl}


\author[0000-0002-1383-0746]{M. Rybak} 
\affiliation{Leiden Observatory, Niels Bohrweg~2, 2333CA Leiden, The Netherlands}

\author[0000-0001-9759-4797]{E. da Cunha}
\affiliation{International Centre for Radio Astronomy Research (ICRAR), University of Western Australia, 35 Stirling Hwy, Crawley, WA 6009, Australia}
\affiliation{ARC Centre of Excellence for All Sky Astrophysics in 3 Dimensions (ASTRO 3D)}
\author[0000-0002-9768-0246]{B. Groves}
\affiliation{International Centre for Radio Astronomy Research (ICRAR), University of Western Australia, 35 Stirling Hwy, Crawley, WA 6009, Australia}
\affiliation{Research School of Astronomy and Astrophysics, Australian National University, Canberra, ACT 2611, Australia}
\author[0000-0001-6586-8845]{J. A. Hodge}
\affiliation{Leiden Observatory, Niels Bohrweg~2, 2333CA Leiden, The Netherlands}
\author[0000-0002-6290-3198]{M. Aravena}
\affiliation{N\'ucleo de Astronom\'ia, Facultad de Ingenier\'ia y Ciencias, Universidad Diego Portales, Av.\,Ej\'ercito 441, Santiago, Chile}
\author[0000-0003-0695-4414]{M. Maseda}
\affiliation{Leiden Observatory, Niels Bohrweg~2, 2333CA Leiden, The Netherlands}
\author[0000-0002-3952-8588]{L. Boogaard}
\affiliation{Leiden Observatory, Niels Bohrweg~2, 2333CA Leiden, The Netherlands}

\author[0000-0002-4153-053X]{D. Berg}
\affiliation{Department of Astronomy, The University of Texas at Austin, 2515 Speedway Blvd Stop C1400, Austin, TX 78712, USA}
\author[0000-0003-3458-2275]{S. Charlot}
\affiliation{Sorbonne Universit\'e, UPMC-CNRS, UMR7095, Institut d'Astrophysique de Paris, F-75014, Paris, France}
\author[0000-0002-2662-8803]{R. Decarli}
\affiliation{INAF --- Osservatorio di Astrofisica e Scienza dello Spazio, via Gobetti 93/3, I-40129, Bologna, Italy}
\author[0000-0001-9714-2758]{D. K. Erb}
\affiliation{The Leonard E. Parker Center for Gravitation, Cosmology and Astrophysics, Department of Physics, University of Wisconsin-Milwaukee, 3135 N Maryland Avenue, Milwaukee, WI 53211, USA}
\author[0000-0002-7524-374X]{E. Nelson}
\affiliation{Department of Astrophysical and Planetary Sciences, University of Colorado, Boulder, CO 80309, USA}
\author{C. Pacifici}
\affiliation{Space Telescope Science Institute, 3700 San Martin Drive, Baltimore, MD 21218, USA}
\author[0000-0002-3418-7251]{K. B. Schmidt}
\affiliation{Leibniz-Institut f\"ur Astrophysik Potsdam (AIP), An der Sternwarte 16, 14482 Potsdam, Germany}
\author[0000-0003-4793-7880]{F. Walter}
\affiliation{Max-Planck Institut f\"ur Astronomie, K\"onigstuhl 17, D-69117, Heidelberg, Germany}
\affiliation{National Radio Astronomy Observatory, Pete V. Domenici Array Science Center, P.O.\,Box O, Socorro 87801, USA}

\author[0000-0002-5027-0135]{A. van der Wel}
\affiliation{Sterrenkundig Observatorium, Department of Physics and Astronomy, Ghent University, Belgium}
\affiliation{Max-Planck Institut f\"ur Astronomie, K\"onigstuhl 17, D-69117, Heidelberg, Germany}

\begin{abstract}
Extreme emission-line galaxies (EELGs) at redshift $z=1-2$ provide a unique view of metal-poor, starburst sources that are the likely drivers of the cosmic reionization at $z\geq6$. However, the molecular gas reservoirs of EELGs - the fuel for their intense star-formation - remain beyond the reach of current facilities.
We present ALMA [\ion{C}{2}] and PdBI CO(2--1) observations of a $z=1.8$, strongly lensed EELG SL2S~0217, a bright Lyman-$\alpha$ emitter with a metallicity 0.05~$Z_\odot$. We obtain a tentative ($\sim$3-4$\sigma$) detection of the [\ion{C}{2}] line and set an upper limit on the [\ion{C}{2}]/SFR ratio of $\leq1\times10^6~L_\odot$/($M_\odot$ yr$^{-1}$), based on the synthesized images and visibility-plane analysis. The CO(2--1) emission is not detected. Photoionization modelling indicates that up to 80\% of the [\ion{C}{2}] emission originates from neutral or molecular gas, although we can not rule out that the gas is fully ionized. The very faint [\ion{C}{2}] emission is in line with both nearby metal-poor dwarfs and high-redshift Lyman~$\alpha$ emitters, and predictions from hydrodynamical simulations. However, the [\ion{C}{2}] line is 30$\times$ fainter than predicted by the De Looze et al. [\ion{C}{2}]-SFR relation for local dwarfs, illustrating the danger of extrapolating locally-calibrated relations to high-redshift, metal-poor galaxies.

\end{abstract}


\section{Introduction}
\label{sec:intro}

The Epoch of Reionization (EoR) is one of the main frontiers of present-day astrophysics. Recent results suggest that the reionization is likely driven by low-metallicity, dwarf ($M_\star\lesssim10^9$ $M_\odot$) galaxies with intense star formation rates; due to their low gas and dust content, a large fraction of UV photons will be able to escape and reionize the neutral intergalactic medium \citep[e.g.,][]{Robertson2010,Robertson2015,Atek2015,Stark2016}. These metal-poor, dwarf galaxies contribute significantly to the cosmic star-forming rate (SFR) at $z\geq3$ \citep[e.g.,][]{Bouwens2009}. Consequently, characterizing the star-forming processes in $z\geq6$ dwarf galaxies presents a crucial step towards understanding the evolution of galaxies at early cosmic times. However, these faint, high-redshift dwarf galaxies remain elusive due to the limitations of current facilities (mainly the \textit{Hubble Space Telescope} and the {\it Spitzer Space Telescope}), along with the fact that for the most distant targets the diagnostic-rich optical emission is shifted to the near-IR regime and thus currently unobservable - a situation that will be soon remedied by the \textit{James Webb Space Telescope}. Consequently, large uncertainties remain on the physical mechanisms of galaxy-led reionization, such as the actual fraction of ionizing photons that escape their interstellar medium (ISM), their star formation efficiency, and feedback processes at play. 

An alternative to studying directly the $z\geq6$ dwarf galaxies is to target the intermediate-redshift ($z\sim2$) extreme emission line galaxies \citep[EELGs;][]{vdWel2011,Atek2011,Maseda2013,Maseda2014, Amorin2015}, which are likely analogues of primordial galaxies at the EoR and can be studied much more efficiently, particularly in emission and absorption lines. EELGs have been identified through extremely high equivalent widths (EWs) of optical emission lines such as [\ion{O}{3}] 5007~{\AA} (with rest-frame EW exceeding 500{\AA}), via excess emission in HST/WFC3 broad-band filters \citep{vdWel2011}, and through HST/WFC3 grism spectroscopy \citep{Maseda2018}. Spectroscopic and photometric follow-up confirm that these are low-mass ($M_\star\sim10^8-10^9\,M_\odot$), low-metallicity ($Z<0.30\,Z_\odot$) dwarf galaxies with high SFRs ($\sim10-100\,M_\odot\rm{yr}^{-1}$), likely undergoing an intense but short-lived burst of star formation \citep[e.g.,][]{Maseda2013,Maseda2014,Masters2014, Tang2019}.

While rest-frame UV and optical studies of high-redshift star-forming dwarfs have primarily targeted the ionized ISM, their intense star formation must be fueled by cold neutral gas. In particular, measuring the cold gas reservoirs of high-redshift dwarfs is crucial for understanding the timescales on which their high-mass star formation can be maintained. However, direct observations of the cold ISM phase in these faint, metal-poor sources are extremely challenging, even at $z\sim1-2$.

Our best bet for probing the molecular gas content of this metal-poor galaxies is the [\ion{C}{2}] 158-$\mu$m line, thanks to its low critical density and large intrinsic brightness. At $z\sim2$, \textit{Herschel} and ALMA Band~9 [\ion{C}{2}] observations have been instrumental in probing the gas content of $M_\star = 10^9-10^{11}$~$M_\odot$ galaxies (e.g., \citealt{Stacey2010, Brisbin2015, Schaerer2015, Zanella2018}).
Crucially, for the EoR sources, the [\ion{C}{2}] line is easily observable with ALMA and has become the chief probe of cool gas in early galaxies (e.g., \citealt{Maiolino2005, Maiolino2015, Knudsen2016, Bradac2017, Matthee2019}; see \citealt{Hodge2020} for a recent review). At the same time, the [\ion{C}{2}] emission in high-redshift galaxies has been explored by a number of cosmological (e.g., \citealt{Olsen2017, Lagache2018}) and zoom-in simulations (e.g., \citealt{Vallini2015, Katz2017, Pallottini2019, Lupi2020}).

In this paper, we attempt to study the neutral ISM in a strongly gravitationally lensed $z\sim2$ SL2S~0217, targeting the [\ion{C}{2}] line with Atacama Large Millimetre/sub-millimetre Array (ALMA), and CO(2--1) with the Plateau de Bure Interferometer (PdBI). Thanks to its large magnification, SL2S~0217 provides a unique opportunity to detect the neutral ISM in a high-redshift EELG with a stellar mass of just $10^8$~$M_\odot$ - 1~dex lower than any previous $z\sim2$ study.

This paper is structured as follows: \S~\ref{sec:obs} presents our ALMA and PdBI observations, data combination and imaging procedures; \S~\ref{sec:results} presents the derivation of the source-plane upper limits on the [\ion{C}{2}] and CO(2--1) luminosities. In \S~\ref{sec:discussion}, we compare our [\ion{C}{2}] non-detection to the expectations photoionization modelling (\S~\ref{subsec:mappings}), various empirical and theoretical [\ion{C}{2}]-SFR relationships (\S~\ref{subsec:comparison_models}) as well as local and high-redshift observations (\S~\ref{subsec:comparison_obs}). Finally, we discuss the fate of the molecular gas in SL2S~0217 (\S~\ref{subsec:mol_gas_fate}) and the prospects of detecting the [\ion{C}{2}] emission from SL2S~0217-like dwarfs in the Epoch of Reionization (\S~\ref{subsec:detection}).

\section{Observations and imaging}
\label{sec:obs}

\subsection{Target description}

SL2S 021737–051329 (henceforth SL2S~0217, J2000 02h 17m 37.237s -05d13m29.7s) is a redshift $z=1.844$ EELG, strongly gravitationally lensed by a $z=0.6459$ elliptical galaxy. SL2S~0217 was serendipitously discovered by \citet[source ID SXDF-iS-170569]{Geach2007}. Thanks to its position under the cusp of the lensing caustic, SL2S~0217 is lensed into a brilliant, 2.5-arcsec long Einstein arc, with a fainter counter-image. The lens system includes a second, doubly-imaged background source SL2S~0217.X at $z=2.29$ \citep{Brammer2012}. 

The stellar content of SL2S~0217 was studied by \citet[B12]{Brammer2012} using HST GRISM spectroscopy and \citet[B18]{Berg2018} using Keck/LRIS optical spectroscopy. Pixellated lens modelling of SL2S~0217 based on the HST data\footnote{see also \citet{Tu2009} and B12 for parametric lens models.} was performed by \citet{Cooray2011}, B18 and \citet{Erb2019}; in the B18 model, the HST F606W continuum is magnified by a factor of $\sim17$, with a source-plane UV half-light radius of $\sim0.35$~kpc. We adopt the source-plane properties listed in Table~\ref{tab:sl2s_properties}, based on the B18 lens model. The $Z=0.05~Z_\odot$ metallicity of SL2S~0217 is typical of $z\sim6$ galaxies with the same stellar mass (c.f. \textsc{FirstLight} simulations, \citealt{Langan2020}), underlining its suitability as a EoR analogue. 

The spatial distribution of the Ly$\alpha$ emission in SL2S~0217 was studied by \citet[E19]{Erb2019} using narrow-band HST imaging. This revealed a 0.6-kpc offset between Ly$\alpha$ and the UV continuum, indicating a varying column density of neutral hydrogen across the source, with the bulk of the Ly$\alpha$ photons escaping along a low column density channel. The long-wavelength spectral energy distribution of SL2S~0217 is only poorly sampled, with a single detection in the Spitzer/MIPS 24-$\mu$m imaging (B12; 7.3--9.1~$\mu$m rest-frame, including the 7.7 and 8.6~$\mu$m PAH bands) which indicates a significant hot-dust and/or PAH emission. In this paper, we extend this comprehensive dataset to the far-infrared and mm-wave regime, targeting the [\ion{C}{2}] 158~$\mu$m line.

We assume a flat $\Lambda$CDM cosmology, with $\Omega_m=0.315$ and $H_0=67.4$ km s$^{-1}$ Mpc$^{-1}$ \citep{Planck2018}. At $z=1.844$, this translates to a luminosity distance $D_L=14460$~Mpc; 1~arcsec corresponds to a physical distance of 8.65~kpc \citep{Wright2006}.

\begin{table}
 \caption{SL2S~0217: intrinsic (de-lensed) properties adapted from \citet[B18]{Berg2018} and \citet[E19]{Erb2019}, with the [\ion{C}{2}] and CO(2--1) upper limits derived assuming the $R\leq3$~kpc aperture from the 1-arcsec taper images (Fig.~\ref{fig:alma_cii}) and a magnification factor $\mu=16$. For the solar metallicity, we adopt the \citet{Asplund2009} value of 12+log(O/H)=8.69. Molecular gas mass $M_\mathrm{H_2}$ is derived from the [\ion{C}{2}] luminosity using the \citet{Madden2020} conversion factor.
   \label{tab:sl2s_properties}}
  \centering
  \begin{tabular}{@{} llcc @{}}
  \hline
 
  & & & Reference\\
  \hline
  $z_\mathrm{spec}$ & & 1.844 & B18 \\
  M$_\star$ & [$M_\odot$] & $(1.8\pm0.4)\times10^8$ & B18 \\
  SFR & [$M_\odot$ yr$^{-1}$] & 23$\pm$2 & B18 \\
  $\mu_\mathrm{UV}$ & & 17.3$\pm$1.2 & B18 \\
  12+log(O/H) & & 7.5 & B18 \\
  $R_\mathrm{1/2}^\mathrm{UV}$ & [kpc$^2$] & 1.0$\times$0.5 & E19 \\
  \hline
      $L_\mathrm{[CII]}$ & [$L_\odot$]& $\leq2.1\times10^7$ (3.2$\sigma$)$^a$& \S~\ref{subsec:cii_limits} \\
    $L'_\mathrm{[CII]}$  & [K km s$^{-1}$ pc$^2$]& $\leq1\times10^8$ (3.2$\sigma$)& \S~\ref{subsec:cii_limits} \\
    $S_\mathrm{160\mu m}$ & mJy & $\leq0.23$ (3$\sigma$)& \S~\ref{subsec:cii_limits} 
   \\

  $L_\mathrm{CO(2-1)}$ & [$L_\odot$]& $\leq1.3\times10^5$ (3$\sigma$) & \S~\ref{subsec:cii_limits} \\
      $L'_\mathrm{CO(2-1)}$  & [K km s$^{-1}$ pc$^2$]& $\leq4\times10^8$ (3$\sigma$) & \S~\ref{subsec:cii_limits} \\
      $M_\mathrm{H_2}^\mathrm{[CII]}$ & [$M_\odot$] & $\leq2.1\times10^9$ (3.2$\sigma$) & \S~\ref{subsec:mol_gas_fate} \\

  \hline
  \multicolumn{4}{l}{$^a$ Based on our tentative image-plane detection, see Fig.~\ref{fig:cii_spectrum}.}
 \end{tabular}
\end{table}

\subsection{ALMA Band 9 observations and imaging}

We combine deep ALMA Band~9 observations of the [\ion{C}{2}] line ($f_0$=1900.539~GHz) and the underlying rest-frame 160-$\mu$m continuum from the ALMA projects \#2016.1.00142.S and \#2016.1.00776.S.

The ALMA programme \#2016.1.00142.S (PI: da Cunha) observations were carried out in two array configurations: C43-1 (2018 July 6) and C43-2 (2018 August 16). The baseline length ranged between 15 and 314 m (C43-1) and 15 to 479 m (C43-2). Forty 12-m antennas were used on both dates. The primary beam FWHM was 9.0~arcsec at 680~GHz. The precipitable water vapour (pwv) ranged between 0.4 and 0.5~mm. The total observing time was 2.7 hours, with a total on-source time of 67~minutes. 

The spectral setup consisted of four spectral windows (SPWs) with 480 channels of 3.906 MHz each, giving a total bandwidth of 2.0 GHz per SPW. The individual SPWs were centered at 667.61, 669.53, 649.91 and 648.03 GHz. 

As the ALMA pipeline products suffered from calibration issues, we calibrated the data manually by completely flagging antennas with high system temperature: C43-1 configuration: DA62, DA44, DA50, DV09, DV24; C43-2: DA62. We used DV07 as the reference antenna due to its low system temperature, good bandpass stability, and central position in the array.

We supplemented these data by observations from the ALMA programme \#2016.1.00776.S (PI: Cooray). These were taken in two array configurations: C43-2 with forty-seven 12-m antennas (2018 October 19) and C43-6 with fifty-one 12-m antennas (2016 October 1 and 14). 
The baseline length ranged between 15 and 3145~m (C43-6) and baselines 15 and 484~m (C43-2), pwv ranged between 0.55 and 0.80~mm. 
The on-source time was 47.0~min for each configuration. The spectral setup consisted of four SPWs configured with 128 15.625-MHz-wide channels (2.0 GHZ bandwidth per SPW), with central frequencies of 664.95, 666.64, 668.34 and 670.03 GHz. For the C43-6 configuration, we manually flagged antennas DA43, DV14, DV17. This reduced the total number of antennas to 48.

For the imaging, we concatenate the visibilities from both ALMA programmes to maximize the S/N; the total on-source time is 160~min. For the C43-6 configuration, we discard all baselines longer than 1000~k$\lambda$, as adding the long baselines significantly reduces the surface brightness sensitivity of the combined dataset. The resulting (u,v)-plane coverage provides sensitivity to spatial scales between 0.21 and 6.2~arcsecs; given the arc length of 2.5~arcsec, we do not expect any structure to be resolved out.

We produce dirty images of the concatenated dataset using natural weighting, at the full angular resolution and using a 0.5 and 1.0-arcsec Gaussian taper. For the [\ion{C}{2}] line, we create several dirty-image cubes using different (u,v)-plane tapers (no taper, 0.5 and 1.0-arcsec taper) and channel width ($\Delta f=$ 50-500~MHz, equivalent to 25-225~km s$^{-1}$). The $\sigma_\mathrm{rms}$ levels in the resulting images match the expected ALMA sensitivity within 10\%.

For the continuum image, we combine all the SPWs, flagging the channels affected by atmospheric lines. The resulting beam sizes and rms sensitivity for the [\ion{C}{2}] and Band~9 continuum imaging are listed in Table~\ref{tab:alma_imaging}. With a sensitivity of $\sigma_\mathrm{rms}=1.3$ mJy/beam over 250~MHz bandwidth (3.7~mJy/beam over 10 km s$^{-1}$), the combined dataset ranks among the deepest ALMA Band~9 observations to-date.

\begin{figure*}
\begin{centering}
\includegraphics[width = 17cm]{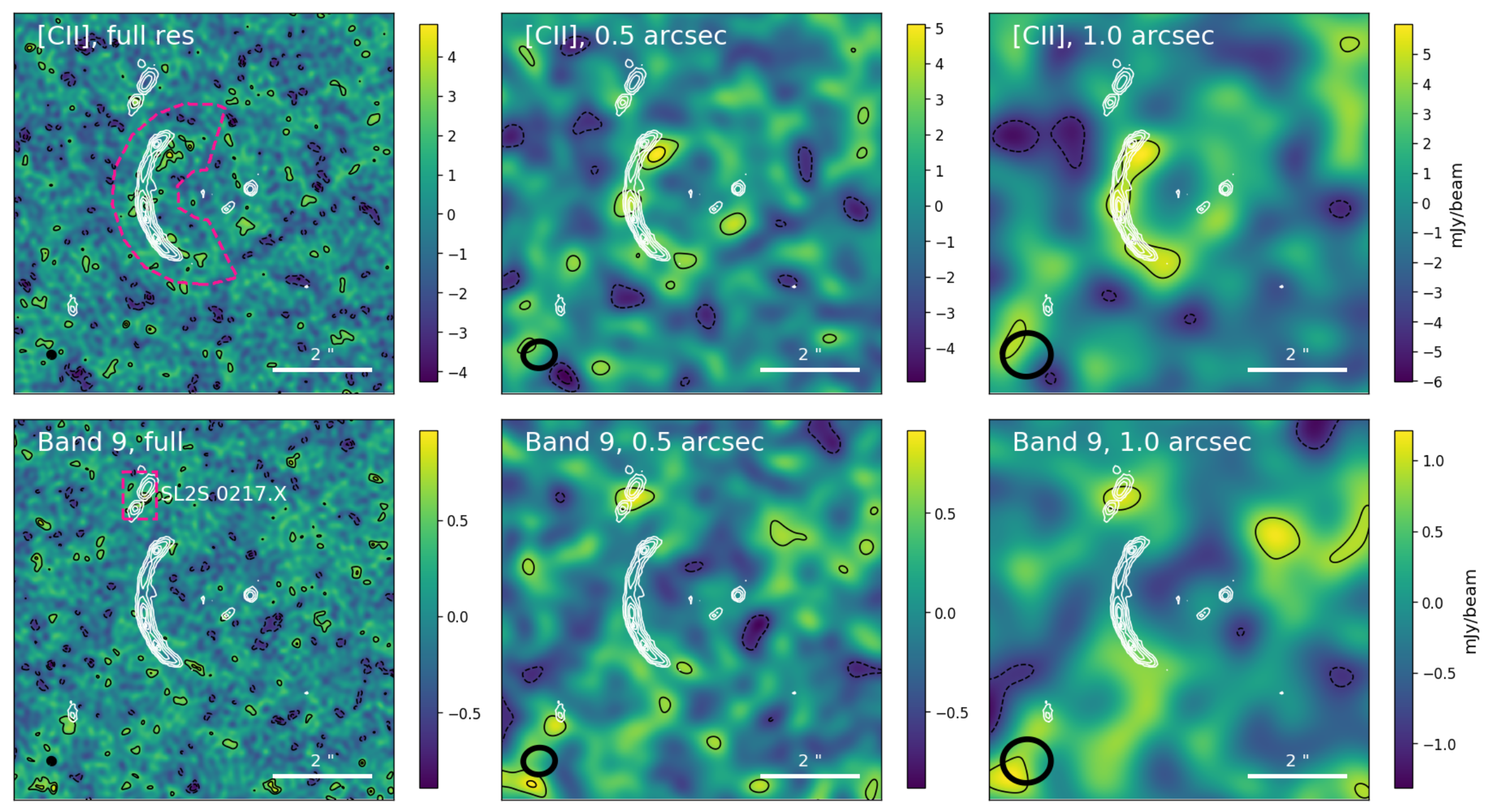}

\end{centering}
\caption{ALMA Band 9 imaging: dirty naturally-weighted images of the [\ion{C}{2}] line (\textit{upper}) and continuum (\textit{lower}) at full resolution and 0.5 and 1.0~arcsec taper, overlayed with the HST F606W imaging (\citealt{Brammer2012}, white contours). The [\ion{C}{2}] emission is integrated over 250~MHz (110 km~s$^{-1}$) bandwidth centered at 667.97~GHz; this part of the spectrum shows the tentative 2-3$\sigma$ extended emission. The black contours start at $\pm2\sigma$ and increase in steps of 1$\sigma$. The HST contours are drawn at 5, 10, 20, 40, 60 and 80\% of the surface brightness maximum. The dashed line (\textit{left}) indicates the $R=3$~kpc aperture used to extract the upper limits (see \S~\ref{subsec:cii_limits}). No continuum emission from the main source is detected; the second lensed source (SL2S~0217.X, $z=2.29$) is tentatively detected at 2-4$\sigma$ significance in each image. \label{fig:alma_cii}}
\end{figure*}

Figures~\ref{fig:alma_cii} presents the resulting synthesised images. We do not find any significant ($\geq5\sigma$) [\ion{C}{2}] or rest-frame 160-$\mu$m continuum emission. However, the [\ion{C}{2}] images show a suggestive 2-3$\sigma$ emission along the main Einstein arc (see discussion below).
In addition, we obtain a tentative (2-4$\sigma$) detection of the rest-frame 140-$\mu$m continuum from the secondary lensed source SL2S 0217.X ($z=2.29$, B12).

The ALMA data were reduced and imaged using the Common Astronomy Software Applications package (\textsc{Casa}, \citealt{McMullin2007}), versions 5.1 and 5.4.

\begin{table}
\caption{ALMA Band~9 synthesized beam size, position angle and rms sensitivity for the [\ion{C}{2}] line (measured at 668~GHz over 250~MHz bandwidth) and the rest-frame 160-$\mu$m continuum as a function of (u,v)-taper. \label{tab:alma_imaging}}
\centering
  \begin{tabular}{@{} c|ccc @{}}
  \hline
taper & beam FWHM (PA) & $\sigma$ ([CII]) & $\sigma$ (cont.)\\
     & [arcsec, deg] & [mJy beam$^{-1}$] & [mJy beam$^{-1}$] \\
  \hline
  full-res & 0.26$\times$0.23 (57) & 1.28 & 0.23 \\
  0.5~arcsec & 0.65$\times$0.57 (-9) & 1.44 & 0.29 \\
  1.0~arcsec & 1.01$\times$0.92 (-9) & 1.91 & 0.44 \\
  \hline
 \end{tabular}
\end{table}

\begin{figure}
\begin{centering}
\includegraphics[width = 8.5cm]{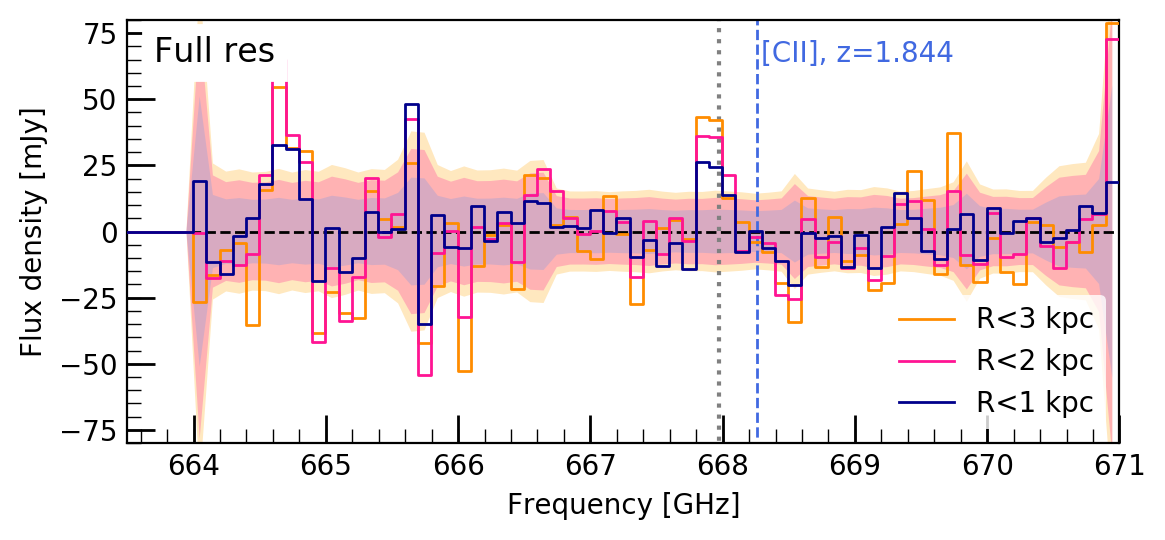}
\includegraphics[width = 8.5cm]{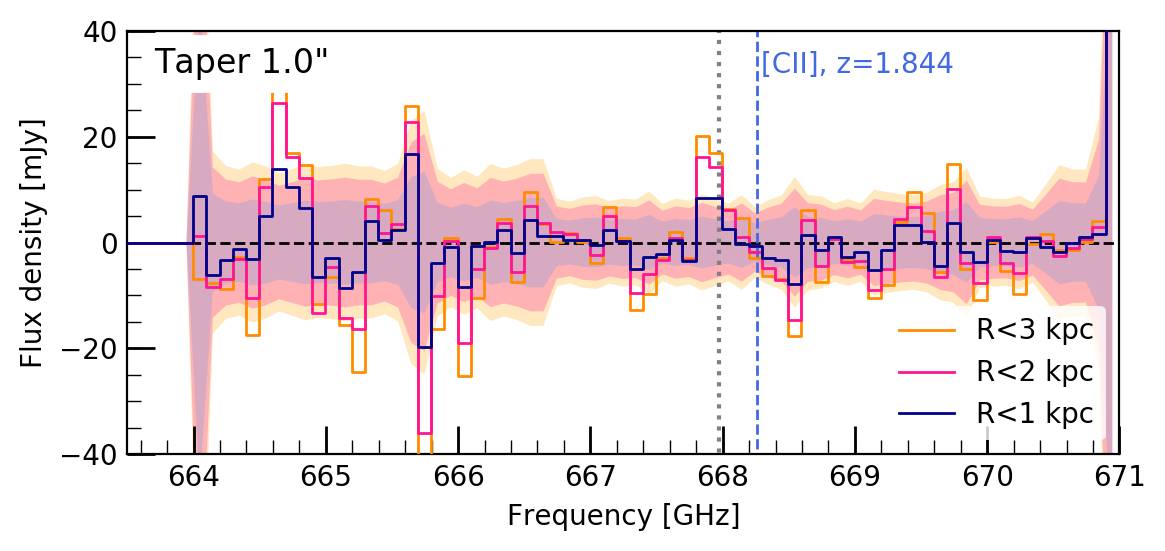}
\end{centering}
\caption{ALMA Band~9 spectrum of SL2S~0217 at 100~MHz (45 km s$^{-1}$) resolution, extracted for the full-resolution and 1.0~arcsec taper images, and the three source-plane apertures. The shaded regions indicate $\sigma_\mathrm{rms}$ for each channel. We obtain a tentative (3-4$\sigma$) [\ion{C}{2}] 158-$\mu$m detection at 667.97~GHz (offset from the systemic velocity by $\sim$50~km\,s$^{-1}$) . \label{fig:cii_spectrum}}
\end{figure}

\subsection{PdBI Band 1 observations and imaging}
\label{subsec:pdbi}

In addition to the ALMA observations of the [\ion{C}{2}] 158-$\mu$m line, we target the CO(2--1) line using the Plateau de Bure Interferometer (PdBI) with the WideX correlator (programme X037, PI: Aravena). The PdBI data were taken between 2013 June 15 and 2013 September 14 in a total of 16 successful tracks. The total on-source time was 29.6~hours (5-antennas equivalent); the total observing time was 40~h. All observations were carried our in the most compact D-configuration, with five 15-meter antennas. The baseline length ranged between 15 and 97~m, resulting in a synthesized beam FWHM = 7~arcsec; the source is thus completely unresolved. The observations covered a frequency range of 79.6 - 83.2~GHz with a spectral resolution of 2~MHz. The resulting rms sensitivity at the expected frequency of the CO(2--1) line is 0.28 mJy beam$^{-1}$ over 100 km s$^{-1}$ bandwidth. The data were reduced using the \textsc{Gildas}/\textsc{Clic} package (\texttt{http://www.iram.fr/IRAMFR/GILDAS}).

\section{Results}
\label{sec:results}

\subsection{[CII] line and 160-$\mu$m continuum}
\label{subsec:cii_limits}

\subsubsection{Searching for signal in the image plane}

To convert the image-plane [\ion{C}{2}] flux to source-plane (intrinsic) flux, we need to account for the gravitational lensing. The magnification of the extended source depends on its surface brightness distribution and the lensing geometry. Although we could simply assume that the [\ion{C}{2}] emission is co-spatial with the HST continuum, resolved observations of high-redshift galaxies have revealed kpc-scale offsets between the rest-frame UV and [\ion{C}{2}] emission (e.g., \citealt{Maiolino2015, Carniani2017}) and very extended [\ion{C}{2}] reservoirs (e.g., \citealt{Carniani2018, Matthee2019, Carniani2020}). If [\ion{C}{2}] is significantly offset from the UV continuum, the [\ion{C}{2}] might be less magnified the UV, increasing the source-plane $L_\mathrm{[CII]}$ upper limit.

We consider three circular source-plane apertures centered on the UV-continuum peak, with a radius $R=$1.0, 2.0 and 3.0~kpc; the latter is comparable to the largest [\ion{C}{2}] reservoirs observed in (much more massive) $z\sim6$ galaxies \citep{Carniani2020}. We project these into the sky-plane using the \citet{Cooray2011} lens model for the lensing galaxy. To maximize the S/N, we consider only the part of the sky-plane aperture corresponding to the main arc (Figure~\ref{fig:alma_cii}); as the arc accounts for 95\% of the total flux. The magnification factor is $\mu$ = 43, 25 and 16 for the $R=$1, 2 and 3-kpc apertures, respectively.

We note that in the synthesis imaging, the area under a dirty beam integrates to zero for large aperture sizes, unlike for a classical point-spread function. Consequently, flux measurements extracted from \emph{dirty} images over extended areas might be biased, particularly in presence of strong positive or negative sidelobes. However, the dirty beams corresponding to the Fig.~\ref{fig:alma_cii} images are well-described by a central Gaussian with only very small sidelobes ($\leq$5\%) within the apertures considered here. Our flux limits should therefore be robust.

Figure~\ref{fig:cii_spectrum} shows the spectra extracted from within these apertures, at 100~MHz resolution, which reveal an excess flux at 667.9~GHz. The excess signal is unlikely to be caused by e.g. continuum contamination or phase errors. At 250~MHz binning (110~km s$^{-1}$), the excess is detected at $3.2-4.3\sigma$ significance, depending on the taper and aperture used. 

No continuum signal is detected at $\geq3\sigma$ significance; we therefore put a conservative 3$\sigma$ upper limit on the rest-frame 160-$\mu$m flux $S_\mathrm{160\mu m}\leq0.23$~mJy, based on the 1.0-arcsec taper images and the $R\leq3$~kpc aperture.

\subsubsection{Searching for signal in the (u,v) plane}

As the spatial filtering by the incomplete (u,v)-plane coverage might decrease the sensitivity to extended emission in the synthesized images, we try to confirm our tentative detection of the [\ion{C}{2}] line in the (u,v)-plane. How bright can the [\ion{C}{2}] emission be to be still consistent with the observed visibilities?

We use the following approach by adapting the (u,v)-plane lens-modelling technique from \citet{Rybak2015a, Rybak2017}. First, we extract the visibility data (real and imaginary parts) at 669.970$\pm$0.125~GHz (same bandwidth as used for the image-plane analysis). The noise on the real/imaginary visibilities is estimated by taking the rms of visibilities for a given baseline for each individual scan. As the noise per polarization might differ, we do not combine the XX and YY polarizations into the Stokes $I$.
We then calculate the expected signal $V(u,v)^\mathrm{model}$ for each visibility as
\begin{equation}
V(u,v)^\mathrm{model} = \sum_{l,m} B(l,m) I(l,m) e^{2\pi i (ul+vm)},
\end{equation}
where $B$ is the primary beam response (approximated by a Gaussian), $I$ the input sky brightness distribution, and $m,l$ the sky-plane coordinates. We consider the following sky-plane [\ion{C}{2}] surface brightness distributions: HST F606W arc (see Fig.~\ref{fig:alma_cii}) and the $R\leq1$, 2 and 3~kpc apertures; the total flux varies between 0 and 50~mJy. We calculate the log-likelihood value $\log L \propto -\sum_i(V(u_i,v_i)^\mathrm{model}-V(u_i,v_i)^\mathrm{data})^2/\sigma(u_i,v_i)^2$. As a control test, we perform the same analysis for a ``line-free'' part of the spectrum (680.000$\pm$0.125~GHz).

\begin{figure}
\begin{centering}
\includegraphics[width = 8.5cm]{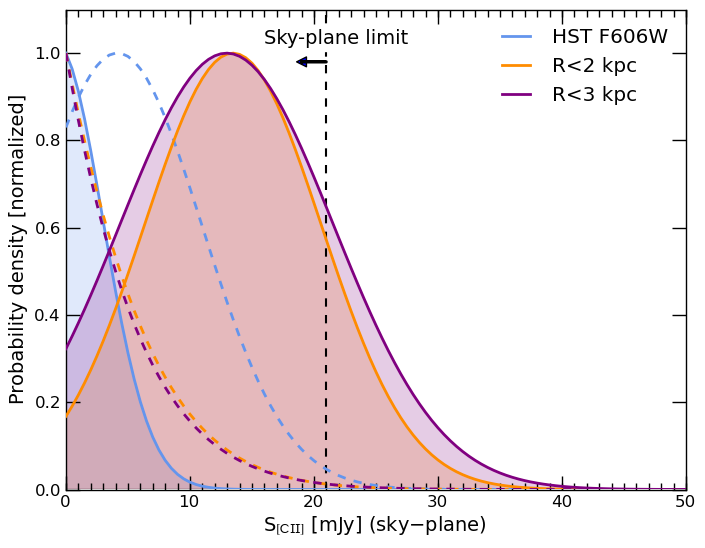}

\end{centering}
\caption{(u,v)-plane analysis of the ALMA [\ion{C}{2}] 158-$\mu$m observations: Probability distribution function (PDF) of the sky-plane [\ion{C}{2}] flux for a 250~MHz bandwidth (see Fig~\ref{fig:alma_cii}), assuming different surface brightness distributions and compared to the sky-plane limit (21 mJy). Colored dashed lines indicate sky-flux PDFs for a ``line-free'' part of the data (i.e., where no signal is expected). Although the [\ion{C}{2}] PDF peaks around 15~mJy for $R\leq2$, 3~kpc apertures, this solution is not strongly preferred over the null hypothesis ($S_\mathrm{[CII]}=0$~mJy) and is below the sky-plane limit. This result does not depend on the noise model used. We therefore adopt an upper limit on $S_\mathrm{[CII]}$=21~mJy based on image-plane analysis.\label{fig:visibilities}}
\end{figure}

Figure~\ref{fig:visibilities} shows the derived probability distribution function (PDF) for $S_\mathrm{[CII]}$. For the $R\leq2$~kpc and $R\leq3$~kpc apertures, the PDF peaks around $S_\mathrm{[CII]}=15$~mJy, whereas for the HST-based aperture, the PDF peaks at 0~mJy. Reassuringly, the PDF for the ``line-free'' part of the spectrum peaks near $S_\mathrm{[CII]}=0$~mJy (dashed lines) for all apertures. Tweaking the noise calculation (e.g., by combining the polarizations, calculating the rms over several scans) does not substantially change the PDFs. This might be indicative of an extended [\ion{C}{2}] emission in SL2S~0217. However, the $S_\mathrm{[CII]}=15$~mJy solution is not strongly preferred over the $S_\mathrm{[CII]}=0$~mJy, our null hypothesis.

Consequently, we adopt a [\ion{C}{2}] upper limit of $S_\mathrm{[CII]}\leq 21$~mJy over 250 MHz (110 km s$^{-1}$) bandwidth, measured for the 1.0-arcsec uv-taper and $R\leq3$~kpc source-plane aperture (3.2$\sigma$ significance). The 1.0-arcsec taper maximizes the surface-brightness sensitivity, while large aperture size accounts for a potentially very extended [\ion{C}{2}] emission. This line flux corresponds to a sky-plane luminosity of $L_\mathrm{[CII]}\leq3.4\times10^8$~$L_\odot$; after de-lensing ($\mu=16$), we set a source-plane upper limit of $L_\mathrm{[CII]}\leq2.1\times10^7$~$L_\odot$. The line width adopted here is comparable to the typical line FWHM of $\sim120$~km$^{-1}$ of $z\sim2$ EELGs \citep{Maseda2014}.

\subsection{CO(2--1) line}

\begin{figure}
\begin{centering}
\includegraphics[width = 8.5cm]{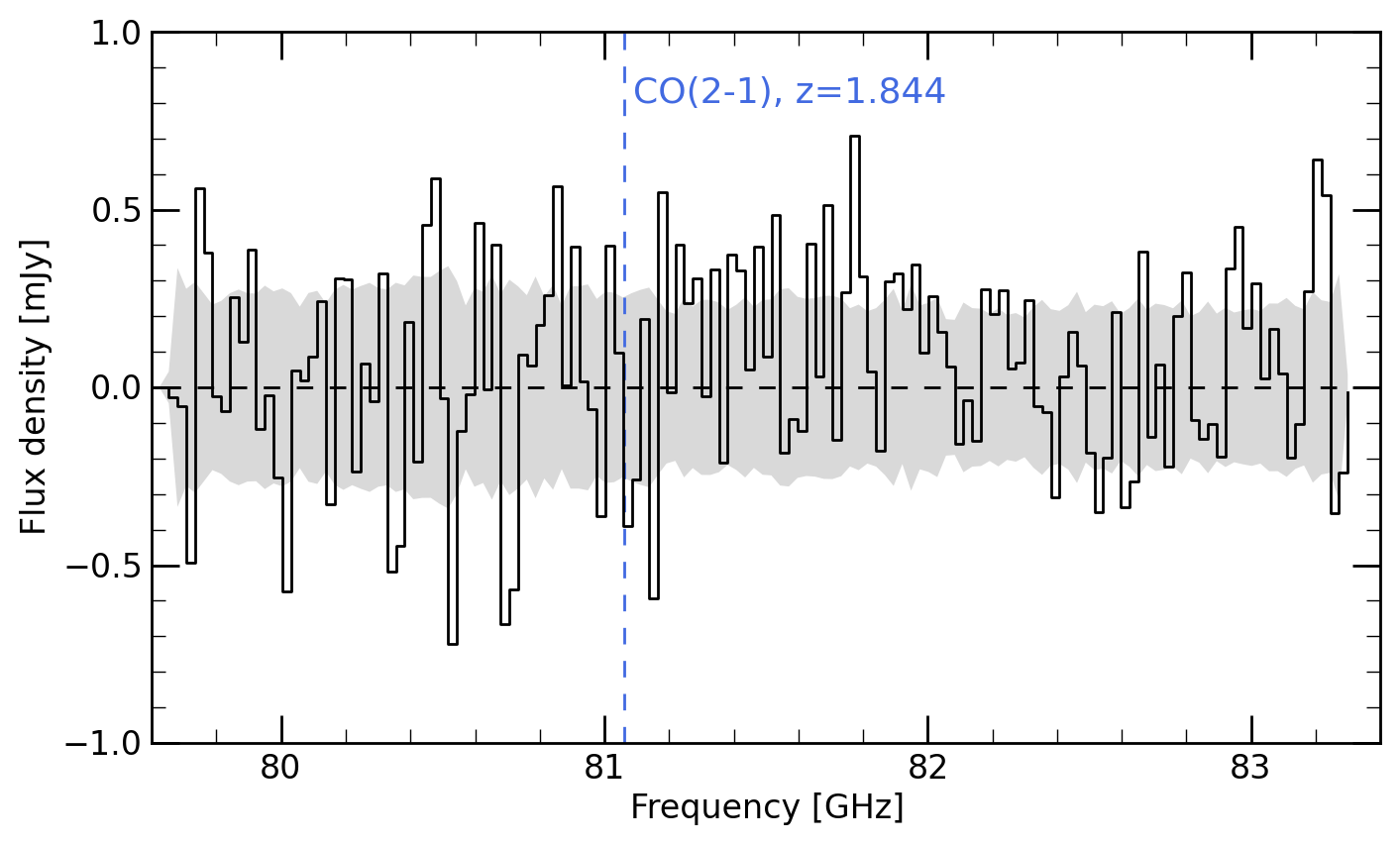}
\end{centering}
\caption{PdBI spectrum of SL2S~0217 with 100~km s$^{-1}$ frequency bins. The shaded region indicates the $\sigma_\mathrm{rms}$ for each channel. We do not detect any significant CO(2--1) emission from SL2S~0217. \label{fig:pdbi}}
\end{figure}

As shown in Figure~\ref{fig:pdbi}, we do not detect any significant CO(2--1) line emission towards SL2S~0217; the PdBI spectrum is consistent with pure noise. Assuming the same aperture and 110~km s$^{-1}$ linewidth as for the [\ion{C}{2}] line (see Figure~\ref{fig:alma_cii}), we set a 3$\sigma$ upper limit on $L_\mathrm{CO(2-1)}\leq1.3\times10^5~L_\odot$, $L'_\mathrm{CO(2-1)}\leq4\times10^8$~K km s$^{-1}$ pc$^{2}$ (source-plane).

\section{Discussion}
\label{sec:discussion}

\subsection{Photoionization modelling}
\label{subsec:mappings}

With an ionization energy of 11.3~eV, the [\ion{C}{2}] 158-$\mu$m emission\footnote{In this section, we explicitly state the wavelengths of individual emission lines for clarity.} can arise from all ISM phases: molecular (H$_2$), neutral (H), and ionized (H$^+$). 
Although studies of nearby galaxies have shown that at $Z\leq0.25~Z_\odot$ almost all [\ion{C}{2}] emission arises from the neutral ISM \citep{Croxall2017, Sutter2019}, the large ionizing flux and limited self-shielding due to low metallicity in SL2S~0217 might cause the ionized component to dominate its $L_\mathrm{[CII]}$ \citep{Ferrara2019}. In SL2S~0217, the semi-forbidden \ion{C}{2}]2325{\AA} emission line and the \ion{C}{2} absorption line detected in the Keck spectra (B18) confirm the presence of at least some C$^+$ in the ionized ISM. But \emph{can} the ionized gas account for the entire [\ion{C}{2}] 158-$\mu$m emission?

We address this question by using photoionization modelling to predict the [\ion{C}{2}] 158-$\mu$m emission from the ionized ISM using the \textsc{Mappings}~V photoionization code \citep{Allen2008,Groves2010}. We focus on the [\ion{C}{2}]~158$\mu$m far-IR line and the rest-frame UV \ion{C}{2}]~2325{\AA} line and the \ion{C}{3}] 1909{\AA} doublet\footnote{Here, we treat the ISM as uniform. However, as shown by E19, the UV spectral slope varies considerably across the source; the \ion{C}{2}], [\ion{C}{2}] and \ion{C}{3}] lines are thus likely not co-spatial on sub-kiloparsec scales., which are all predicted by \textsc{Mappings}~V}. The photoionization modelling provides a lower limit on $L_\mathrm{[CII]}$, as the contribution of [\ion{C}{2}]~158~$\mu$m from photon-dominated regions (PDRs) \citep[e.g.,][]{Hollenbach1997} is not included in the \textsc{Mappings} models. Due to the lack of constraints on the PDR phase (just our [\ion{C}{2}] 158-$\mu$m and a very weak CO(2--1) upper limits, we refrain from modelling the neutral/molecular ISM.

The semi-forbidden \ion{C}{3}] 1909{\AA} doublet originates in the \ion{H}{2} regions (e.g., \citealt{Jaskot2016, Kewley2019, Vallini2020}), and - thanks to its brightness - is a convenient tracer of ionized gas in the EoR \citep{Stark2015, Stark2017, Zitrin2015}; the \ion{C}{2}] 2325{\AA} line traces the outer layers of \ion{H}{2} regions, which also emit the [\ion{C}{2}] 158~$\mu$m line.
Based on our upper limits and B18 spectroscopy, the observed ratios are [\ion{C}{2}] 158~$\mu$m / \ion{C}{2}]~2325{\AA}$\leq$ 4.7, [\ion{C}{2}] 158~$\mu$m / \ion{C}{3}]~1909{\AA} $\leq$ 0.8 and \ion{C}{3}]~1909{\AA} / \ion{C}{2}]~2325{\AA} = 6.2$\pm$0.5 (in W m$^{-2}$ units). Using \ion{C}{2}] instead of any of the other UV lines detected in the Keck spectrum by B18 avoids systematic uncertainties related to the carbon abundance and ionization state. 

To set the ionizing radiation field, we use the \textsc{Starburst99} stellar population models\footnote{For the case considered here, the predicted line intensities differ only marginally ($\leq0.1$~dex) between the different stellar model libraries. See \citealt{DAgostino2019} for a detailed comparison.} \citep{Leitherer1999,Leitherer2010}, including also the effects of stellar rotation on the emitted spectra of massive stars, as described in \cite{Levesque2012}. 

\begin{figure}
\begin{centering}
\includegraphics[height = 6cm]{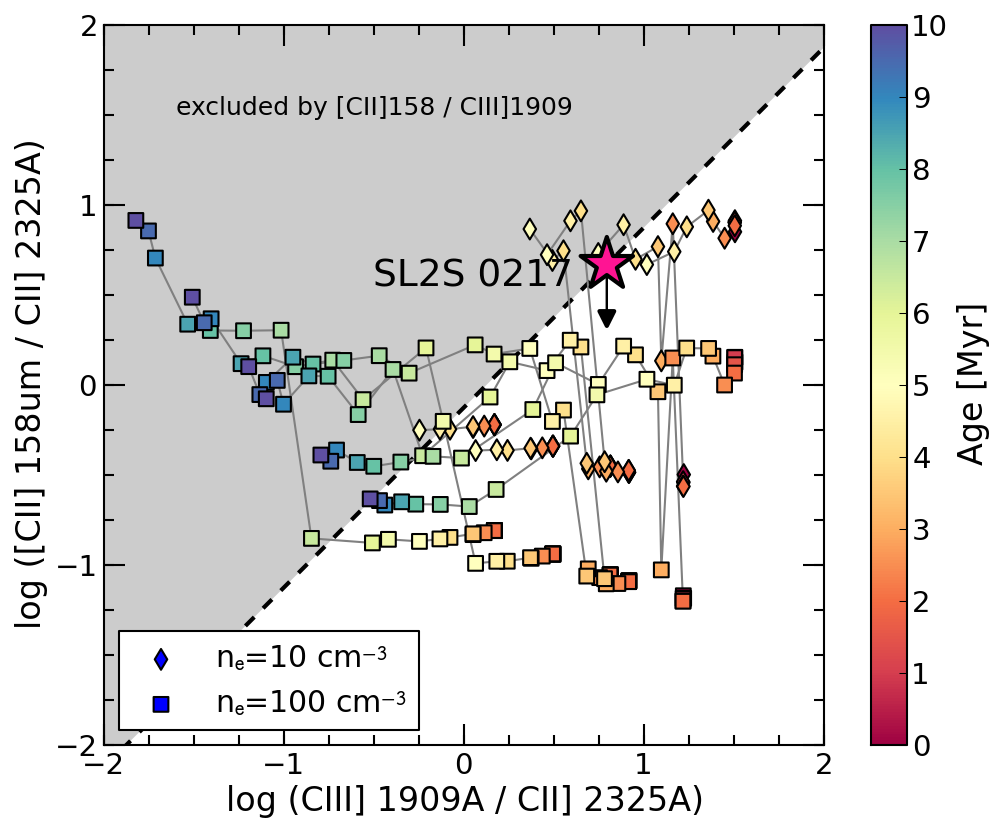}
\includegraphics[height = 6cm]{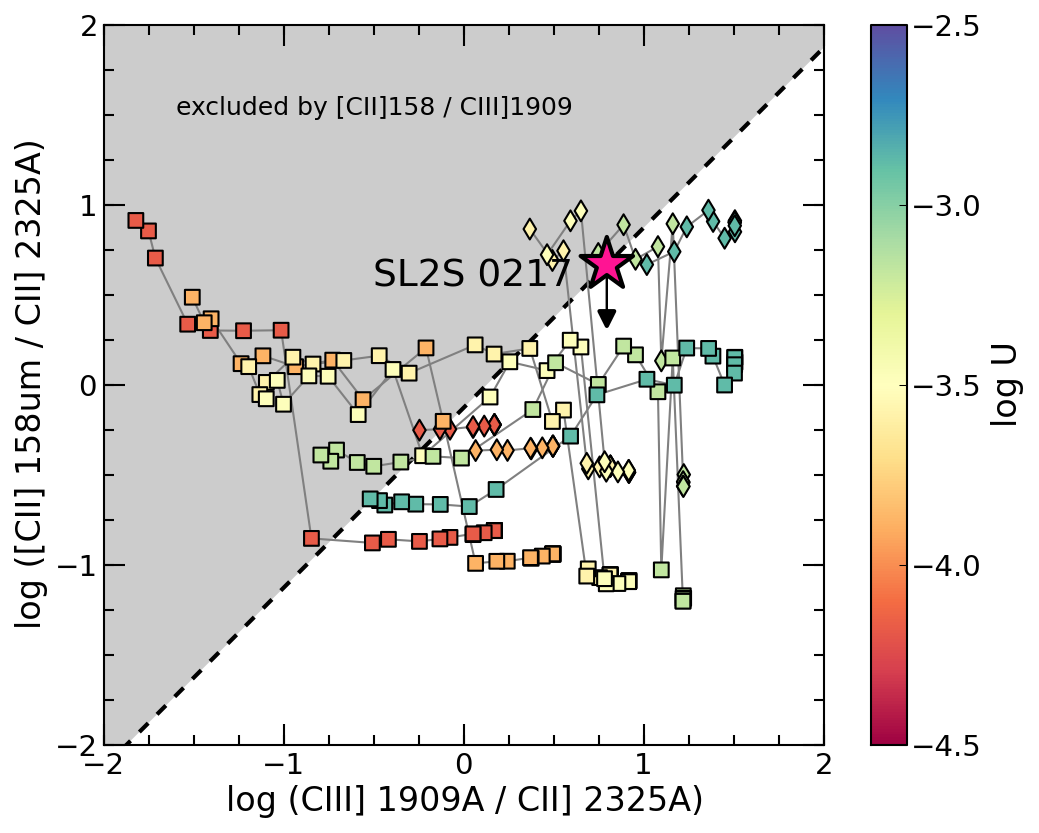}

\end{centering}
\caption{\textsc{Mappings} modelling of C$^+$/C$^{2+}$ emission lines from the ionized ISM as a function of the starburst age (\textit{upper}) and the ionization parameter $U$ (\textit{lower}). We show the ratios of [\ion{C}{2}] 158$\mu$m, \ion{C}{2}]~2325{\AA} and \ion{C}{3}]~1909{\AA} luminosities for an instantaneous star-formation burst, $Z=0.05~Z_\odot$, $n_e=$10 and 100~cm$^{-3}$. The grey lines connect models with constant $U$ and increasing age. The shaded area is excluded by the [\ion{C}{2}]~158~$\mu$m / \ion{C}{3}]~1909{\AA} ratio. The upper limit on $L_\mathrm{[CII]}$ is consistent with a young ($<$5~Myr) starburst over a wide range $U$ and allows for a substantial [\ion{C}{2}] emission from the neutral ISM.
 \label{fig:mappings}}
\end{figure}

Following the B18 modelling of the UV spectrum, we set the model metallicity to $0.05\,Z_\odot$ and assume an instantaneous starburst. We explore models with electron densities between $n_e$=10 and 100 cm$^{-3}$, since the \ion{C}{2}]~2325{\AA}/ [\ion{C}{2}]~158$\mu$m flux ratio is very sensitive to density due to the low critical density of the [CII]~158$\mu$m line in ionised gas ($n_{e}$=50~cm$^3$). We vary the ionization parameter $U$ (defined as the ratio of the $\geq13.6$~eV photon number density and the gas density $n$) and the age of the ionizing star cluster (in steps of 0.5~Myr). We adopt a one-dimensional plane-parallel geometry as appropriate for the expected geometry in SL2S~0217: a screen of ionized ISM on front of an ionizing cluster. 
Figure~\ref{fig:mappings} shows the predicted [\ion{C}{2}] 158~$\mu$m / \ion{C}{2}]~2325{\AA} and \ion{C}{3}]~1909{\AA} / \ion{C}{2}]~2325{\AA} ratios for each realization.
 
The \ion{C}{3}]~1909{\AA}/\ion{C}{2}] 2325~{\AA} ratio depends both on the ionisation parameter (impacting both the gas temperature and ionization state) and the age of the starburst (affecting the relative number of C$^+$ ionizing photons, $>$24.4~eV, to C$^0$ ionizing photons, $>$11.3~eV). See \citet{Jaskot2016, Kewley2019} for other examples, and the color axes in Fig.~\ref{fig:mappings}.

As the [\ion{C}{2}] 158~$\mu$m and \ion{C}{2}] 2325~{\AA} lines arise only from a single species and ionisation state, their ratio is purely dependent upon the electron density and temperature, which drive the relative collisional rates. Because of the low critical density for [\ion{C}{2}] 158~$\mu$m ($n_e\simeq50$ cm$^{-3}$) and high excitation energy of the \ion{C}{2}] 2325~{\AA} line, the [\ion{C}{2}] 158~$\mu$m / \ion{C}{2}] 2325~{\AA} ratio depends strongly on both electron density and temperature.

We find that, for $n_e=100$\,cm$^{-3}$ (the value assumed by B18), \textsc{Mappings} predicts $\log(L_\mathrm{[CII]158\mu m}^\mathrm{ion}/L_\mathrm{CII]2325\AA}) \sim -0.5 \pm 0.5$ (Figure~\ref{fig:mappings}). For $n_e=10$\,cm$^{-3}$, the ratio is typically 0.5 dex higher, as the critical density for [\ion{C}{2}] 158$\mu$m is lower than for \ion{C}{2}]2325{\AA}. 
Therefore, we conservatively assume that $\log(L_\mathrm{[CII]158\mu m}^\mathrm{ion}/L_\mathrm{CII]2325\AA})$ = 0$\pm$1. Given the \ion{C}{2}]2325{\AA} flux of $1.11\times10^{-17}$~erg s$^{-1}$ cm$^{-2}$ (flux of the brightest line in the \ion{C}{2}] triplet, measured by B18; this is the line predicted by \textsc{Mappings}), we estimate an on-sky [\ion{C}{2}] 158$\mu$m line luminosity $L_\mathrm{[CII]158\mu m}^\mathrm{ion}\sim7.3\times10^7\,L_\sun$. 

This estimate is approximately 5$\times$ lower than our [\ion{C}{2}]~158 $\mu$m upper limit ($L_\mathrm{[CII]158\mu m}^\mathrm{sky}\leq3.4\times10^8\,L_\sun$) although the lack of $n_e$ constraints implies a $\sim$1~dex uncertainty in the predicted flux ratio. 

For most of the $n_e$, $U$ models considered here, our [\ion{C}{2}] 158 $\mu$m upper limit is higher than the contribution from the ionized ISM; the neutral/molecular ISM can account for up to 80\% of the [\ion{C}{2}] 158~$\mu$m line flux, similar to the $z=0$ low-metallicity galaxies \citep{Croxall2017, Cormier2019}. However, in the low-density, high-ionization regime ($n_e=10$~cm$^{-3}$, log $U\geq-3.5$), the predicted [\ion{C}{2}] 158-$\mu$m luminosity exceeds our upper limits. In such case, the ISM in SL2S~0217 will be fully ionized.

\subsection{Comparison with empirical models and simulations}
\label{subsec:comparison_models}

We now compare SL2S~0217 to predictions for [\ion{C}{2}] luminosity from various empirical models and hydrodynamical simulations. 
Thanks to the stream of high-redshift [\ion{C}{2}] detections from ALMA \citep{Hodge2020}, the [\ion{C}{2}] emission from high-redshift galaxies has been extensively studied by a number of high-resolution hydrodynamical simulations. Our tentative [\ion{C}{2}] detection in a $Z=0.05~Z_\odot$ dwarf allows us to validate these models in the poorly explored high-SFR, low-metallicity regime.

\begin{figure*}
\begin{centering}
\includegraphics[height = 7.2cm]{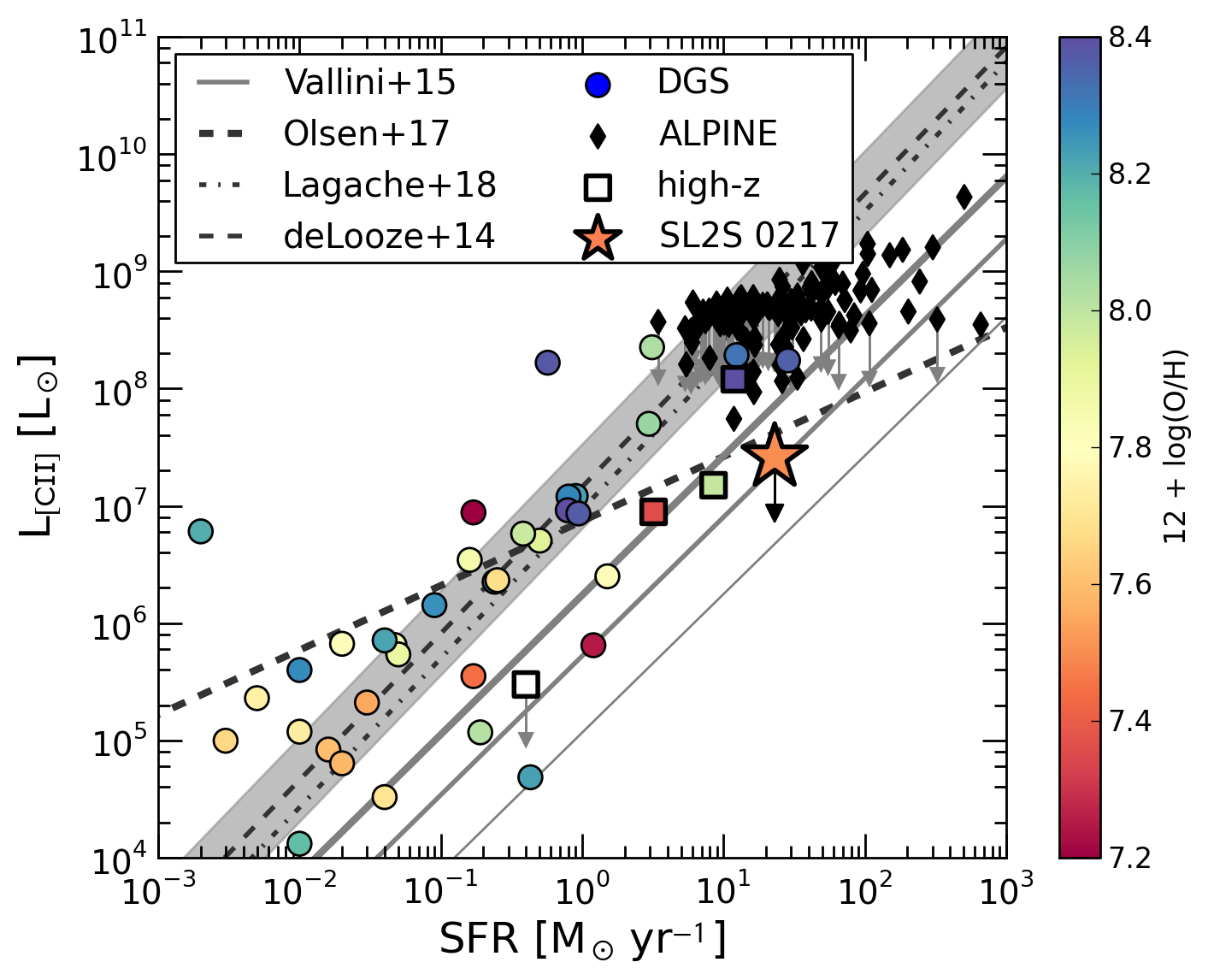}
\includegraphics[height = 7.2cm]{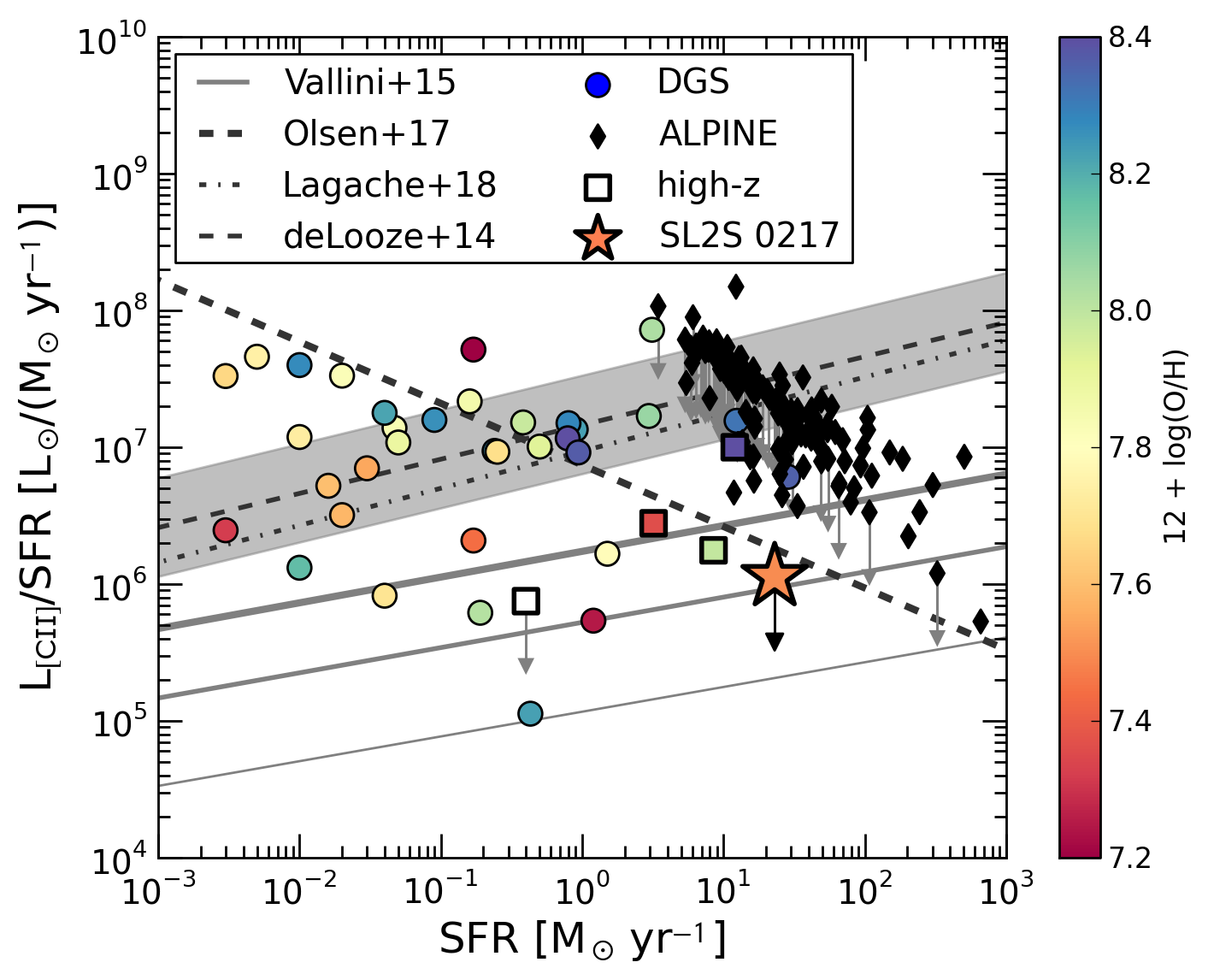}

\end{centering}
\caption{[\ion{C}{2}] luminosity (\textit{left}) and the [\ion{C}{2}]/SFR ratio versus SFR (\textit{right}) in SL2S~0217, compared to the nearby Dwarf Galaxy Survey \citep{Cormier2015, Cormier2019} and \textsc{Alpine} samples \citep{Schaerer2020}, and $z=2-7$ observations of low-mass galaxies \citep{Schaerer2015, Knudsen2016, Bradac2017}, colored by 12 + log (O/H). For the \textsc{Alpine} sources, we use the publicly-available SED-based SFR estimates from \citet{Faisst2019}. The different lines indicate different models from Table~\ref{tab:relations}: \citet{Vallini2015} for $Z=0.025$ (thinnest line), 0.05, and 0.1~$Z_\odot$ (thickest); \citet{deLooze2014} (with 1$\sigma$ scatter indicated by shading); \citet{Olsen2017} and \citet{Lagache2018}. We do not show the \citet{Herrera2015} relation, as it generally predicts higher $L_\mathrm{[CII]}$ than \citet{deLooze2014} over the SFR range studied here. The [\ion{C}{2}]/SFR ratio in SL2S~0217 is at least 30$\times$ lower than predicted by the \citet{deLooze2014} relation. \label{fig:cii_sfr}}
\end{figure*}

\begin{figure}
\begin{centering}
\includegraphics[height = 6.5cm]{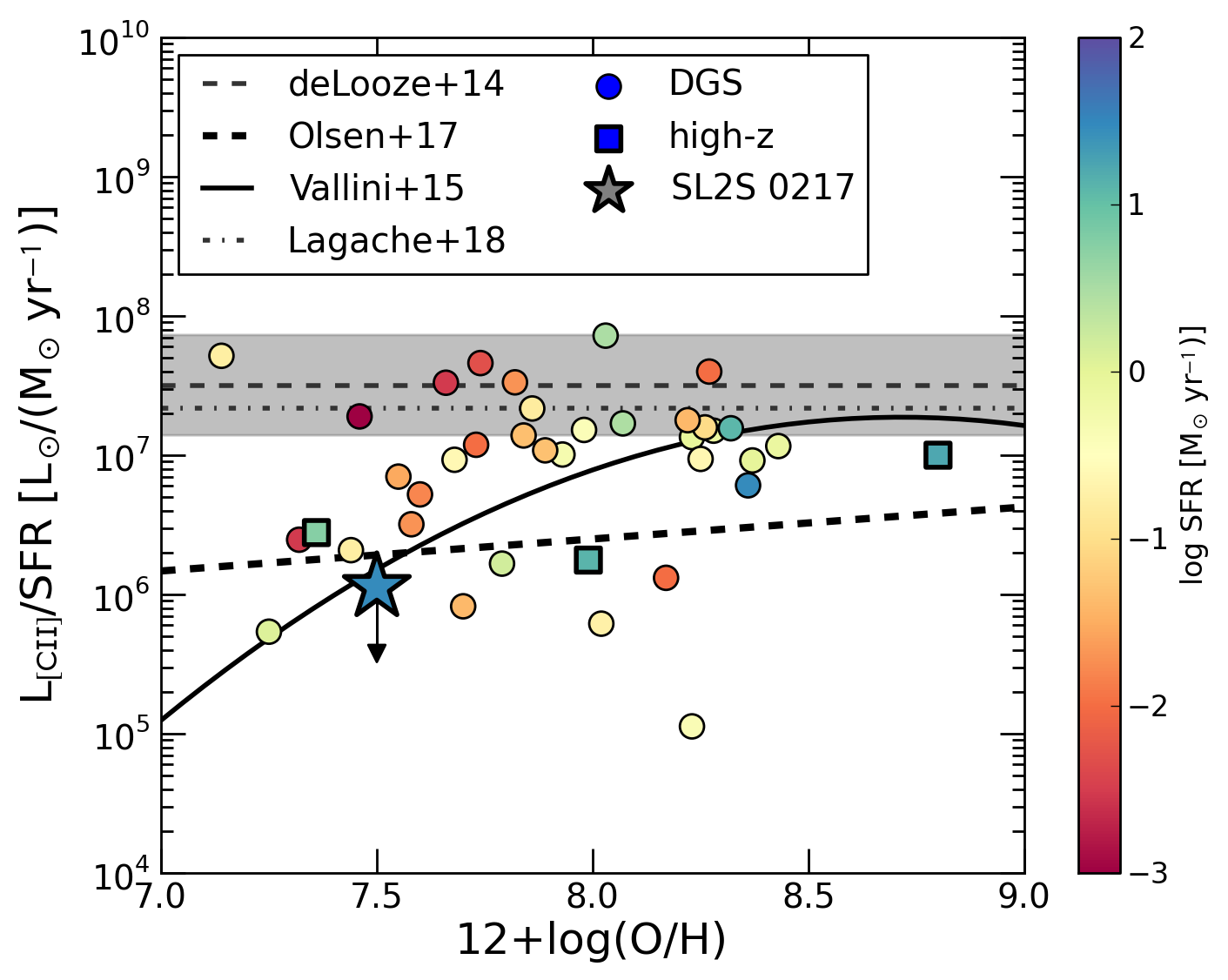}

\end{centering}
\caption{[\ion{C}{2}]/SFR ratio as a function of metallicity: comparison of SL2S~0217 to the \citet{Cormier2015}, \citet{Knudsen2016} and \citet{Bradac2017} sources (color-coded by SFR), and the different relations from \S~\ref{subsec:comparison_models}, all for SL2S~0217-like SFR=23~$M_\odot$ yr$^{-1}$. The [\ion{C}{2}]/SFR ratio in SL2S~0217 is consistent with the metallicity-dependent relations of \citet{Vallini2015} and \citet{Olsen2017}, but much lower than predicted by \citet{deLooze2014} and \citet{Lagache2018}.
 \label{fig:logOH_cii_sfr}}
\end{figure}

Namely, we consider the following [\ion{C}{2}]-SFR relations: the empirical \citet{deLooze2014} relation for nearby low-metallicity dwarf galaxies, the \citet{Herrera2015} calibration for nearby star-forming galaxies, as well as predictions from simulations of \citet{Vallini2015, Olsen2017} and \citet{Lagache2018}. Table~\ref{tab:relations} lists the respective [\ion{C}{2}]-SFR relations and the predicted source-plane [\ion{C}{2}]/SFR ratio for SL2S~0217. Figures \ref{fig:cii_sfr} and \ref{fig:logOH_cii_sfr} compare the $L_\mathrm{[CII]}$ and $L_\mathrm{[CII]}$/SFR in SL2S~0217 with relations from Table~\ref{tab:relations}, and low- and high-redshift observations (see below) as a function of SFR and metallicity.

Our upper limit is in agreement with the metallicity-dependent relations of \citet{Vallini2015} and \citet{Olsen2017}. On the other hand, the [\ion{C}{2}] emission in SL2S~0217 is at least 30$\times$ fainter than expected from the locally-calibrated \citet{deLooze2014}, a $\geq$4$\sigma$ tension. The \citet{Herrera2015} relation predicts even higher $L_\mathrm{[CII]}$/SFR with a smaller scatter ($\sim0.2$~dex); this is likely a consequence of a relatively high metallicity of their sample (12 + log(O/H)=7.7-8.8).

Finally, \citet{Lagache2018} underpredict the SL2S~0217 [\ion{C}{2}] luminosity by a factor of 20; however, given the 0.5~dex 1$\sigma$ scatter of the \citet{Lagache2018} trend, our upper limit is still consistent with it at 2.5$\sigma$ level. We note that the \citet{Lagache2018} simulations contain a significant number of galaxies with SFR, metallicity and $L_\mathrm{[CII]}$ consistent with SL2S~0217.

The strong dependence of the $L_\mathrm{[CII]}$/SFR ratio on average gas metallicity is supported by the high-resolution simulations of high-$z$ dwarf galaxies by \citet{Lupi2020}: for $Z\leq0.1~Z_\odot$, they find $L_\mathrm{[CII]}$/SFR$\leq10^6$~$L_\odot$/($M_\odot$ yr$^{-1}$), consistent with our upper limit.

Finally, we consider the analytical model of \citet{Ferrara2019}, who investigated low [\ion{C}{2}] luminosities of some of the high-$z$ UV-selected sources. In the low-metallicity, high-$\Sigma_\mathrm{SFR}$ regime of SL2S~0217, the [\ion{C}{2}] surface density $\Sigma_\mathrm{[CII]}$ is given by:

\begin{equation}
\Sigma_\mathrm{[CII]}=1.8\times10^7 \left(\frac{n_\mathrm{gas}}{n_{500}} \right) \ln\left(\frac{U}{0.001}\right),
\label{eq:ferrara}
\end{equation}

where $n_{500}=500$~cm$^{-3}$.
For SL2S~0217, with an ionization parameter $\log U = -1.5$ (B18) and assuming that [\ion{C}{2}] is co-spatial with the UV continuum, Eq.~(\ref{eq:ferrara}) matches our $\Sigma_\mathrm{[CII]}$ upper limit for $n_\mathrm{gas}\simeq 150$~cm$^{-3}$ ($n_\mathrm{gas}$ decreases further if [\ion{C}{2}] is more extended than the UV continuum). As the Ferrara et al. model directly sets $U\propto\Sigma_\mathrm{SFR}/\Sigma_\mathrm{gas}^2$, to reconcile the high $U$ and $\langle \Sigma_\mathrm{SFR} \rangle$, SL2S~0217 has to be an extreme starburst, with a gas surface density $\Sigma_\mathrm{gas}\simeq3.5\times10^8$~$M_\odot$ kpc$^{-2}$. This roughly agrees with the expectation from the E19 gas column density estimate $N_\mathrm{HI}\simeq10^{21.7}$~cm$^{-3}$, which predicts $\Sigma_\mathrm{gas}= N_\mathrm{HI} \times m_\mathrm{proton} \simeq 8\times10^7$~$M_\odot$ kpc$^{-2}$.

\begin{table*}
 \caption{[\ion{C}{2}]-SFR relations referenced in this work, with the expected [\ion{C}{2}]/SFR ratio for SL2S~2017, given the parameters from Table~\ref{tab:sl2s_properties}. The [\ion{C}{2}]/SFR ratio in SL2S~0217 is $\leq1\times10^6$~$L_\odot$/($M_\odot$~yr$^{-1}$). \label{tab:relations}}
  \centering
  \begin{tabular}{@{} llll @{}}
  \hline
  Reference & [\ion{C}{2}]-SFR & $L_\mathrm{[CII]}$/SFR & Note \\
  & & [$L_\odot$/($M_\odot$~yr$^{-1}$)] & \\
  \hline
 \citet{deLooze2014} & $\log L_\mathrm{[CII]} = 1.25\log (\mathrm{SFR}) + 7.16$ & $3\times10^7$ & Local dwarf galaxies \citep{Cormier2015}\\
\citet{Herrera2015} & $\log L_\mathrm{[CII]}= 0.97\log (\mathrm{SFR}) + 7.66$ & $4\times10^7$ & Local star-forming galaxies\\
 \citet{Vallini2015} & $\log L_\mathrm{[CII]} = 7.0 + 1.2\log(\mathrm{SFR}) + 0.021\log(Z)+$ & $9\times10^5$ & Zoom-in simulations, $z=6.6$ \\
 & \hspace{1cm}$+ 0.012 \log(\mathrm{SFR}) \log(Z) - 0.74 \log^2(Z)$ & & $Z=0.05 - 1~Z_\odot$, SFR = 1-100~$M_\odot$ yr$^{-1}$\\
 \citet{Olsen2017}$^a$ & $\log L_\mathrm{[CII]} = 7.17 + 0.55 \log \mathrm{SFR} + 0.23 \log Z$ & $1.8\times10^6$ & Zoom-in simulations, $z\sim6$,  \\
& & & $M_\star=(6-80)\times10^8$~$M_\odot$,\\
& & &  SFR = $3-20~M_\odot$~yr$^{-1}$, $Z=0.16-0.45~Z_\odot$\\
 \citet{Lagache2018}$^b$ & $\log L_\mathrm{[CII]} = 7.1+\left(1.4-0.07 z\right) \log(\mathrm{SFR})-$ & $2.2\times10^7$& Semi-analytic models, $z=4-7$ \\
  & $-0.07z$ & &$Z=0.004-4.4$~$Z_\odot$ \\
  
  \hline
\multicolumn{4}{l}{$^a$ not correcting for the CMB temperature effects which are negligible at $z\sim2$ (c.f. \citealt{Olsen2018} erratum).}\\
\multicolumn{4}{l}{$^b$ no metallicity dependence, weak evolution with redshift $z$.}

 \end{tabular}

\end{table*}

\subsection{Comparison with low-metallicity galaxies near and far}
\label{subsec:comparison_obs}

In the present-day Universe, SL2S~0217 can be directly compared to galaxies from the Dwarf Galaxy Survey (DGS) \citep{Madden2013}, which span $Z$ = 0.02 - 1.0~$Z_\odot$ and SFR = 0.0005 - 25 $M_\odot$ yr$^{-1}$; 48 galaxies from the DGS sample were targeted with \textit{Herschel} PACS far-IR spectroscopy by \citet{Cormier2015}. Although the DGS sample shows a broad [\ion{C}{2}]/SFR correlation, the [\ion{C}{2}]/SFR ratio varies between $10^{5.0}$ and $10^{9.5}$~$L_\odot/(M_\odot \mathrm{yr}^{-1})$ (Figure~\ref{fig:cii_sfr}). Four DGS sources have [\ion{C}{2}]/SFR ratios similar or lower than SL2S~0217. In particular, SBS~0335-052 - which matches SL2S~0217 in terms of the rest-frame UV line ratios and the overall SED (E18) - has [\ion{C}{2}]/SFR $\simeq5.5\times10^5$~$L_\odot$/($M_\odot$ yr$^{-1}$), directly comparable to our upper limits.

At $z=1-2$, the bulk of [\ion{C}{2}] and CO molecular gas studies have focused on relatively massive galaxies with $M_\star\geq10^10~M_\odot$ (e.g., \citealt{Stacey2010, Brisbin2015, Zanella2018}). The only exception is the strongly lensed $M_\star \simeq 2.5\times10^9 M_\odot$ galaxy detected in [\ion{C}{2}] and CO(3--2) by \citet{Schaerer2015}; however, it is still $\sim10\times$ more massive and metal-enriched than SL2S~0217. The [\ion{C}{2}]/SFR ratio measured by \citet{Schaerer2015} is $10^7~L_\odot$ / ($M_\odot$~yr$^{-1}$), 1~dex higher than in SL2S~0217. 

At $z\geq6$, two $z=6-7$ EELGs lensed by galaxy clusters have been detected in the [\ion{C}{2}] emission by \citet[who also obtained a strong upper limit on another source]{Knudsen2016} and \citet{Bradac2017}. Similar to SL2S~0217, all three sources have [\ion{C}{2}]/SFR $\leq3\times10^6$, significantly lower that the locally-calibrated \citet{deLooze2014} relation (Figure~\ref{fig:cii_sfr}). While the \citet{Knudsen2016} [\ion{C}{2}]-detected source has a relatively low SFR and high stellar mass ($M_\star\simeq3\times10^9$~$M_\odot$), the \citet{Bradac2017} source is very similar to SL2S~0217, with SFR$\simeq9$~$M_\odot$ yr$^{-1}$, $M_\star\simeq10^8$~$M_\odot$ and $Z=0.2$~$Z_\odot$.

At higher stellar masses ($M_\star=10^{9}-10^{11}~M_\odot$), the \textsc{Alpine} ALMA large programme \citep{LeFevre2019, Faisst2019, Bethermin2020} has recently observed the [\ion{C}{2}] line in 189 $z=4-6$ galaxies with SFR = $1-100$~$M_\odot$ yr$^{-1}$ \citep{Schaerer2020}. The \textsc{Alpine} sources are generally consistent with the \citet{deLooze2014} relation and have higher [\ion{C}{2}]/SFR than SL2S~0217, likely due to their presumably higher metallicity (23 out of 118 \textsc{Alpine} sources are detected in the rest-frame 160-$\mu$m continuum indicating substantial dust masses, \citealt{Bethermin2020}).

\begin{figure}
\begin{centering}
\includegraphics[width = 8.3cm]{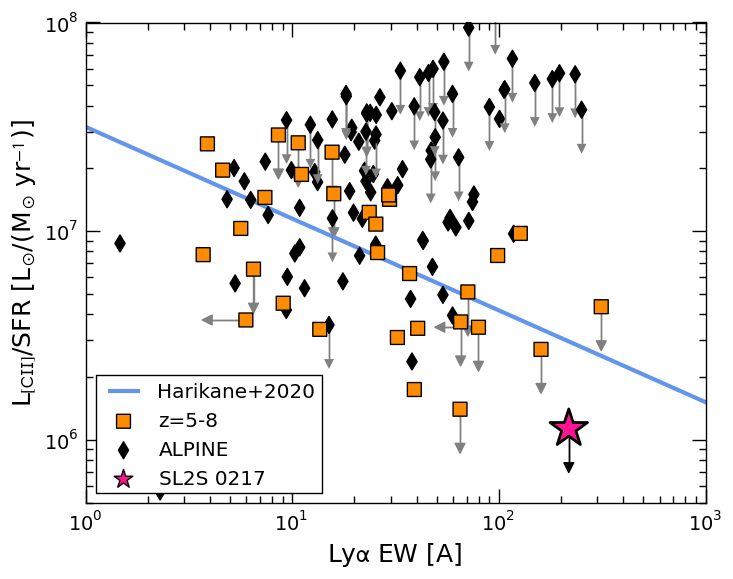}

\end{centering}
\caption{Upper limits on the [\ion{C}{2}]/SFR ratio in SL2S~0217 versus the Ly$\alpha$ equivalent width (E19), compared to the \citet{Harikane2018} and \citet{Matthee2019} compilations of $z=5-7$ Ly$\alpha$ emitters and the $z=4.4-5.9$ \textsc{Alpine} survey \citet{Schaerer2020}. SL2S~0217 is a factor of 3 below the empirical relation of \citet{Harikane2020}. \label{fig:ew_cii_sfr}}
\end{figure}

Finally, Figure~\ref{fig:ew_cii_sfr} compares the [\ion{C}{2}] luminosity and [\ion{C}{2}]/SFR ratio in SL2S~0217 to the compilation of $z=5-7$ Ly$\alpha$ emitters \citep{Harikane2018, Matthee2019}, and the \textsc{Alpine} survey \citep{Schaerer2020}. The Ly$\alpha$ EW correlates closely with the Ly$\alpha$ photons escape fraction (e.g., \citealt{Verhamme2017, Harikane2018, Matthee2019}); the sources with high Ly$\alpha$ EW are thus expected to be depleted in neutral ISM. For SL2S~0217, E19 derived a photometric EW = 218$\pm$12 {\AA}. Compared to the $z\geq5$ Ly~$\alpha$ emitters, SL2S~0217 has a low [\ion{C}{2}]/SFR ratio and is a factor of $\sim3$ below the \citet{Harikane2020} empirical Ly$\alpha$~EW - [\ion{C}{2}]/SFR relation. The low [\ion{C}{2}]/SFR ratio in SL2S~0217 might be driven by its relatively compact size: its source-plane $R_{1/2}^\mathrm{UV}\sim$0.35kpc, compared to $\sim$1~kpc radii for $z\sim$6 sources \citep{Carniani2018, Carniani2020, Matthee2019} which might increase the ionized ISM fraction, suppressing the [\ion{C}{2}] emission.

\subsection{The fate of molecular gas in SL2S~0217}
\label{subsec:mol_gas_fate}
Given the upper limits on the [\ion{C}{2}] and CO(2--1) luminosity, what can we say about the state of the molecular gas in SL2S~0217? Does SL2S~0217 still contain a substantial molecular gas reservoir, or is it at the very end of its starburst phase, with the ISM depleted and ionized? We briefly discuss three facets of this problem: i) limits on the molecular gas mass, ii) the possibility of the ISM being fully ionized and iii) photoevaporation of molecular gas.

First, adopting the conservative magnification factor $\mu=16$, the CO(2--1) upper limit translates to a (very weak) 3$\sigma$ upper limit $M_\mathrm{H_2}\leq 2\times10^9~M_\odot$, where we assume $r_{2/1} = 0.6$ \citep{Cormier2014} and a Galactic $\alpha_\mathrm{CO} = 4.4M_\odot$/(K km s$^{-1}$ pc$^{2}$). However, studies of nearby low-metallicity dwarfs (e.g., \citealt{Schruba2012, Hunt2015, Shi2016}) found $\alpha_\mathrm{CO}$ up to 3~dex higher than the Galactic value, significantly weakening our $M_\mathrm{gas}$ upper limit.

Alternatively, we can estimate the molecular gas mass $M_\mathrm{H_2}$ from the [\ion{C}{2}] luminosity, as originally proposed by \citet{Zanella2018}. Due to the very low metallicity of SL2S~0217, rather than using the \citet{Zanella2018} relation (see below), we use the recently-published relation of \citet{Madden2020} calibrated on the DGS sample:
\begin{equation}
M_\mathrm{H_2}^\mathrm{[CII]} \, (\mathrm{M20}) = 132\times L_\mathrm{[CII]}^{0.97},
\end{equation}. 
which for SL2S~0217 yields $M_\mathrm{H_2}\leq2.1\times10^9$~$M_\odot$. 
Alternatively, using the \citet{Zanella2018} relation calibrated on more massive main-sequence galaxies:
\begin{equation}
M_\mathrm{H_2}^\mathrm{[CII]} \, (\mathrm{Z18})= 30\times L_\mathrm{[CII]},
\end{equation}
we obtain $M_\mathrm{H_2}\leq0.8\times10^9$~$M_\odot$; although the extrapolation of the \citet{Zanella2018} relation to the low-$Z$ regime might not be straightforward. Both $M_\mathrm{gas}$ estimates imply a high gas mass fraction $f_\mathrm{gas}=M_\mathrm{gas}/(M_\mathrm{gas}+M_\star)\simeq 0.8$. This is consistent with high gas fractions and dynamical mass estimates for $z=1-2$ EELGs \citep{Maseda2014} and simulations of high-redshift dwarf galaxies \citep{Ceverino2018}.

A gas mass of $\sim10^8$~$M_\odot$ is also supported by the expectations from the \citet{Ferrara2019} model and the column-density estimates from E19 (see \S~\ref{subsec:comparison_models}).

A gas mass of $10^8$~$M_\odot$ is further supported by the high $\Sigma_\mathrm{SFR}$/$U$ ratio in SL2S~0217 (see \S~\ref{subsec:comparison_models}). With SFR=23~$M_\odot$ yr$^{-1}$, SL2S~0217 will deplete its gas in $\sim$10~Myr (likely much faster, as some gas might be expelled from the system). This suggests that SL2S~0217 is in a final stage of an intense starburst.

Second, as indicated in \S~\ref{subsec:mappings}, a substantial fraction of the gas reservoir might be ionized. Namely, the maximum column density proposed by E19 ($\Sigma_\mathrm{HI}\leq10^{21.7}$~cm$^{-2}$) is comparable to the ionized gas column depth of $\sim10^{21.3}$~cm$^{-2}$ predicted by \citet{Ferrara2019} for SL2S~0217-like ionization parameter and metallicity. The fraction of the ionized gas might be further increased due to additional ionizing sources. In particular, the unusually strong \ion{He}{2} emission in SL2S~0217 cannot be reproduced by standard photoionization models (B18) and requires additional ionization source, such as radiative shocks \citep{Allen2008, Plat2019} or high-mass X-ray binaries as seen in, e.g., a nearby dwarf I~Zw~18 (\citealt{Lebouteiller2017, Schaerer2019}, but c.f.\,\citealt{Plat2019}). This extra ionization would increase the contribution of the ionized ISM to the [\ion{C}{2}] luminosity compared to our photoionization models; on the other hand, strong FUV fields might suppress the [\ion{C}{2}] emission by ionizing C$^+$ into C$^{2+}$ (e.g., \citealt{Langer2015}).

Third, the combination of the strong FUV fields and the lack of ISM self-shielding at low metallicities can lead to a rapid ionization and photoevaporation of the molecular clouds on 1-10~Myr timescales, thus further suppressing the [\ion{C}{2}] emission \citep{Vallini2017}.

We note that in the very near future, the warm neutral or molecular gas might be traced at mid-infrared wavelengths via the PAHs or rotational-vibrational H$_2$ emission using the \textit{James Webb Space Telescope}. Alternatively, the gas mass in SL2S~0217 can be constrained kinematically (e.g., \citealt{Calistro2018}), using the integrated-field spectroscopy of the bright \ion{C}{3}]~1909{\AA} line (Maseda et al., in prep.). 

\subsection{Detectability of [CII] emission from metal-poor dwarfs at $z\geq6$}
\label{subsec:detection}

What are the prospects of detecting the [\ion{C}{2}] 158-$\mu$m emission in SL2S~0217 if it was at $z\sim6$? Assuming $L_\mathrm{[CII]}=2.6\times10^7$~$L_\odot$, achieving a spatially unresolved 5$\sigma$ detection at $z=6$ would require $\sigma_\mathrm{rms}\sim50$~$\mu$Jy over 100 km~s$^{-1}$ bandwidth, which corresponds to $\sim$10~hours of ALMA on-source time. For comparison, the deepest ALMA Band~6 observations to-date, delivered by the \textsc{Aspecs} \citep{Walter2016, Gonzalez2020} and \textsc{Almacal} \citep{Oteo2016} projects and the \citet{Fujimoto2016} compilation, reach sensitivities of $\sigma_\mathrm{rms}\sim100$~$\mu$Jy over the same bandwidth.

Another promising way of studying the population of galaxies at high redshift is the line-intensity mapping (e.g. \citealt{Gong2012, Silva2015}). The feasibility of the [\ion{C}{2}]-intensity mapping measurements depends critically on the [\ion{C}{2}] luminosity function at $z\geq6$. A recent [\ion{C}{2}]-intensity mapping feasibility study of \citet{Yue2019} considered several [\ion{C}{2}] luminosity function models. The detectability of the [\ion{C}{2}] power spectrum is the highest for the \citet{deLooze2014} dwarf-galaxies relation, and lowest for the \citet{Vallini2015} relation; the predictions for the expected power-spectrum and shot-noise signal from the two models differ by a factor of $\sim$30. Consequently, if the low [\ion{C}{2}]/SFR ratio in SL2S~0217 can be taken to validate the \citet{Vallini2015} and \citet{Olsen2017} [\ion{C}{2}] luminosity models, the $z\geq6$ [\ion{C}{2}] intensity mapping signal will fall below the detection threshold of the potential ALMA or single-dish intensity-mapping experiments (e.g., a 1000-hour programme on the CCAT-p telescope, \citealt{Stacey2018}). Our analysis highlights the need to properly account for the metallicity evolution of galaxies and its impact on the [\ion{C}{2}] emission.

\section{Summary and Conclusions}
\label{sec:conclusion}

We have presented deep ALMA and PdBI observations of the [\ion{C}{2}] 158-$\mu$m and CO(2--1) emission in the $z=1.844$ strongly lensed metal-poor dwarf galaxy SL2S~0217. In one of the deepest ALMA Band~9 observations to-date, we obtain a tentative 3-4$\sigma$ detection of the [\ion{C}{2}] line. No CO(2--1) emission is detected in the PdBI observations.

Our main conclusions are:
\begin{itemize}
\item We report a tentative (3-4$\sigma$) detection of the [\ion{C}{2}] line based on the image-plane spectra; the source-plane luminosity is $L_\mathrm{[CII]}\leq 2.1 \times 10^7$~$L_\odot$, derived from 1.0-arcsec taper imaging, $R\leq$3~kpc aperture and a 250~MHz bandwidth. We do not find strong evidence for [\ion{C}{2}] emission in the (u,v)-plane. The rest-frame 160-$\mu$m continuum and the CO(2--1) line are not detected. Our tentative [\ion{C}{2}] detection extends the molecular gas studies at $z\sim2$ by 1~dex in stellar mass, down to the $M_\star\simeq10^8$~$M_\odot$ regime.

\item The upper limit on the $L_\mathrm{[CII]}$/SFR ratio in SL2S~0217 is $2\times10^6$~$L_\odot$/($M_\odot$ yr$^{-1}$), within the range spanned by nearby dwarf galaxies \citep{Cormier2015}, as well as $z\geq5$ lensed dwarf EELGs and Ly$\alpha$ emitters.

\item The [\ion{C}{2}]/ SFR ratio in SL2S~0217 is consistent with the \citet{Vallini2015} and \citet{Olsen2017} simulations-based models. However, the [\ion{C}{2}]/SFR ratio in SL2S~0217 is $30\times$ lower than predicted by the locally-calibrated \citet{deLooze2014} relation.

\item We use \textsc{Mappings} photoionization modelling to predict the fraction of [\ion{C}{2}] emission arising in the ionized ISM. We find that our upper limit leaves room for a significant (up to 80\%) contribution from the PDRs; however, we can not exclude that the ISM is fully ionized, particularly for low electron density. The contribution from the ionized ISM might be boosted due to extra ionizing radiation from e.g., shocks or X-ray binaries. Future ALMA and JWST observations of PDR tracers will allow us to directly probe the neutral/molecular ISM.

\item The tentative [\ion{C}{2}] detection and the high $\Sigma_\mathrm{SFR}$ and ionization parameter suggest a gas mass of few $10^8$~$M_\odot$. In this scenario, SL2S~0217 will be in a late phase of an extreme starburst and deplete its gas reservoir in $<$10~Myr. More robust constraints on the gas mass might be obtained from spatially-resolved spectroscopic observations of the bright rest-frame FUV lines.

\item If SL2S~0217 is representative of the $z\geq6$ low-mass galaxies, these will be within reach of deep-field ALMA observations (even without gravitational lensing). On the other hand, a strong dependence of [\ion{C}{2}]/SFR ratio on metallicity might strongly suppress the [\ion{C}{2}] signal in line-intensity mapping experiments. 
\end{itemize}

The large discrepancy between the [\ion{C}{2}]/SFR ratio in SL2S~0217 and the widely-used \citet{deLooze2014} relation for present-day dwarf galaxies highlights the limitations of applying locally established relations to high redshift. At the same time, the good agreement between our tentative [\ion{C}{2}] detection, high-redshift observations, and simulations confirms that - though challenging to study - SL2S~0217 remains a powerful analogue of the sources from the Epoch of Reionization.

\acknowledgments

We thank the anonymous referee for their careful and constructive review of this manuscript which improved the content and clarity of this paper.

This paper makes use of the following ALMA data: ADS/JAO.ALMA \#2016.1.00142.S and \#2016.1.00776.S. ALMA is a partnership of ESO (representing its member states), NSF (USA) and NINS (Japan), together with NRC (Canada), MOST and ASIAA (Taiwan), and KASI (Republic of Korea), in cooperation with the Republic of Chile. The Joint ALMA Observatory is operated by ESO, AUI/NRAO and NAOJ.
This work is based on observations carried out under project number X037 with the IRAM PdBI Interferometer. IRAM is supported by INSU/CNRS (France), MPG (Germany) and IGN (Spain).
We acknowledge assistance from Allegro, the European ALMA Regional Center node in the Netherlands.

MR and JAH acknowledge support of the VIDI research programme with project number 639.042.611, which is (partly) financed by the Netherlands Organisation for Scientific Research (NWO). 
MR acknowledges support from the Leids Kerkhoven-Bosscha Fonds, subsidy number~19.2.075. 
EdC gratefully acknowledges the Australian Research Council as the recipient of a Future Fellowship (project FT150100079) and the ARC Centre of Excellence for All Sky Astrophysics in 3 Dimensions (ASTRO 3D; project CE170100013).
MA has been supported by the grant “CONICYT + PCI +
INSTITUTO MAX PLANCK DE ASTRONOMIA MPG190030”
and “CONICYT+PCI+REDES 190194.”
DKE is supported by the US National Science Foundation through the Faculty Early Career Development (CAREER) Program, grant AST-1255591.
CP is supported by the Canadian Space Agency under a contract with NRC Herzberg Astronomy and Astrophysics.

\facilities{ALMA, IRAM:Interferometer}

\clearpage

\bibliography{SL2Sbibliography}{}
\bibliographystyle{aasjournal}


\end{document}